# Mechanical, Optical and Thermoelectric Properties of Janus BiTeCl Monolayer

Poonam Chauhan, Jaspreet Singh, and Ashok Kumar[*]

*Department of Physics, School of Basic Sciences, Central University of Punjab, Bathinda, 151401, India*

(April 18, 2022)

\* Corresponding author: ashokphy@cup.edu.in

**Abstract**

We report mechanical, optical and thermoelectric properties of recently fabricated Janus BiTeCl monolayer using density functional and semi-classical Boltzmann transport theory. Janus BiTeCl monolayer exhibits a direct bandgap, high carrier mobility (~$10^3$ cm$^2$V$^{-1}$s$^{-1}$) and high optical absorption in the UV-visible region. The mechanical behavior of the Janus BiTeCl monolayer is nearly isotropic having an ideal tensile strength ~ 15 GPa. The higher value of the Gruneisen parameter ($\gamma$), a low value of phonon group velocity ($v_g$), and very little phonon scattering time ($\tau_p$) lead to low lattice thermal conductivity (1.46 W/mK) of Janus BiTeCl monolayer. The combined effect of thermal conductivity and electronic transport coefficients of Janus BiTeCl monolayer results in the figure of merit (ZT) in the range of 0.43-0.75 at 300-500 K. Our results suggest Janus BiTeCl monolayer be a potential candidate for optoelectronic and moderate temperature thermoelectric applications.

## 1. Introduction

In recent years, the new derivative of two-dimensional (2D) materials namely 'Janus monolayers' have attracted considerable research attention due to their distinct properties as compared to the traditional 2D compounds[1]. Janus monolayers contain two different types of atoms on the opposite sides of the 2D surface. 2D Janus structures such as MoSSe have been successfully fabricated in the experiments [2, 3]. The recent theoretical investigations of various 2D Janus monolayers such as MXX' (M= Mo, W ; X/X' = S, Se, Te) [4-7], $Al_2XX'$ (X/X' = O, S, Se and Te) [8], $In_2XO$ (X= S, Se and Te) [9] and group-III monochalcogenides [10] have also been reported.

2D Janus transition metal dichalcogenides (TMDs) are semiconductors with an energy bandgap ranging from 0.6 eV to 1.97 eV [11-14]. The strong optical absorption of light occurs in the UV-visible region of these monolayers [15]. The electronic structure and optical absorption spectra of these monolayers can be further tuned with applied mechanical strains [4]. The tunable finite bandgap and strong visible light absorption in Janus TMDs monolayers make them potential candidates in a wide range of applications including energy harvesting [16], electromechanical systems[16], and optoelectronic devices [11, 15, 17].

Very recently, bismuth-based 2D Janus monolayers BiTeBr and BiTeCl are successfully fabricated [18], which exhibit trigonal structure with the $P6_3mc$ space group. Their crystal structure is asymmetric as Bi atomic planes are stacked in between the atomic planes of Te and Cl/Br atoms. The atomic arrangements lead to the large induced electric field along the stacking direction of these monolayers. Also, Janus BiXY (X= S, Se, Te, and Y = F, Cl, I) monolayers have been investigated by using the first-principles density functional theory [19]. These monolayers are reported to be thermodynamically stable and their bandgap ranges from infrared to visible region of optical spectrum [19]. These monolayers satisfy the Born and Haung [20] criteria that indicate their mechanical stability.

2D layered materials also show promises in thermoelectric applications [21]. The efficiency of thermoelectric materials is measured in terms of the figure of merit ($ZT=S^2\sigma T/\kappa_e+\kappa_l$), where S is Seebeck coefficient, $\sigma$ is electrical conductivity, T is temperature, $\kappa_e$ is electrical thermal conductivity, and $\kappa_l$ is lattice thermal conductivity. The term $S^2\sigma$ in the numerator is known as power factor (PF). The key requirement of thermoelectric is low lattice thermal conductivity i.e. soft and anharmonic [22]. ZT of thermoelectric materials can be enhanced by decreasing the thermal conductivity and increasing the power factor.

2D layered materials are reported to exhibit excellent thermoelectric properties e.g. gamma-graphyne, TMDs [4], phosphorene [23] and SnSe [24] show enhanced ZT value than their bulk counterparts. Janus WSSe and WSTe have ZT values ~0.7 at room temperature (300K) [6]. Also, the thermoelectric performance of 2D Janus BiTeBr (ZT ~0.7) is calculated to be better than that of its bulk counterpart (ZT~0.1) [25]. Also, very recently the thermoelectric performance of AsSBr (0.21-0.53) [26] and $AsSbC_3$ (0.99-0.95) [27] has been investigated theoretically which suggests these materials as potential candidates for thermoelectric applications.

Motivated by the recent experimental synthesis of Bi-based 2D Janus materials and their exciting properties, we have systematically investigated the mechanical, optical and thermoelectric properties of the 2D Janus BiTeCl monolayer. Our first-principles calculations show high carrier mobility of BiTeCl monolayer. The mechanical properties are quantified in terms of ideal strength, ultimate strain, Young's modulus and Poisson ratio. The optical properties are analyzed in terms of optical absorbance and electron energy loss spectra. Furthermore, the thermoelectric performance of Janus BiTeCl has been determined in terms of ZT by utilizing the electronic transport coefficients and lattice thermal conductivity ($k_l$).

## 2. Computational Methodology

First-principles density functional theory based calculations are performed using Quantum Espresso Package [28, 29]. The exchange-correlation effects are included by generalized gradient approximation (GGA) through Perdew-Burke-Ernzehof (PBE) functional [30]. Considering the high atomic masses of Bi, Te and Cl, the spin-orbit coupling (SOC) effects are also included in the calculations. The plane wave cut-off of 80 Ry and the Monkhorst-pack meshes of 24×24×1 is used for sampling of Brillouin zone. In the self-consistent calculations for geometric relaxations, the convergence threshold is set as $10^{-4}$ eV and $10^{-3}$ eV/Å for total energy and atomic forces, respectively. A vacuum space of 14 Å is used to annihilate the interactions between the adjacent layers of Janus BiTeCl along the perpendicular direction. For optical calculations, we have used the dense optical mesh of 40x40x1 with the optical broadening of 0.1 eV. The $G_0W_0$ approximation and BSE (Bethe-Salpeter equation) method has been used to investigate the optical properties as implemented in YAMBO code [31]. The convergence is assured by taking 200 bands with k-point mesh of 20x20x1 for GW and BSE calculations. In the phonon calculations, the phonon spectrum is obtained using supercell of size 6x6x1. The electronic transport coefficients are obtained using 150x150x1 k-points with BoltzTraP code [32]. The lattice thermal conductivity is calculated by using phono3py with relaxation time approximation [33]. The mesh point of 40x40x1 is used for lattice thermal conductivity part. The electronic and lattice parameters are normalized by $L_z/d^*$, where $L_z$ is the length of the unit cell along z-direction and $d^*$ is the effective thickness of 2D material. The effective thickness ($d^*$) of the BiTeCl monolayer is taken as a summation of van der Waal radii of surface atoms (Te, Cl) and the thickness of the system.

## 3. Results and Discussion

The relaxed structure of the BiTeCl monolayer is shown in Figure 1. The optimized lattice constant of the hexagonal unit cell is 4.31 Å which shows an excellent agreement with the previously reported value (4.31Å) [34]. The cohesive energy is calculated to be 3.53 eV/atom which is higher than the other related 2D materials such as BiXY (X = S, Se, Te; Y = Cl, Br, I) [19]. The phonon spectrum is free from the imaginary mode of frequencies over the Brillouin zone (Figure S1, ESI) and, indicating that the BiTeCl monolayer is dynamically stable. To further explore the stability of Janus BiTeCl monolayer, we have performed MD simulations at 300K and 700K, which show small fluctuations in energy and temperature around the constant level w.r.t. time steps (Figure S2, ESI), with no noticeable distortion in final structure.

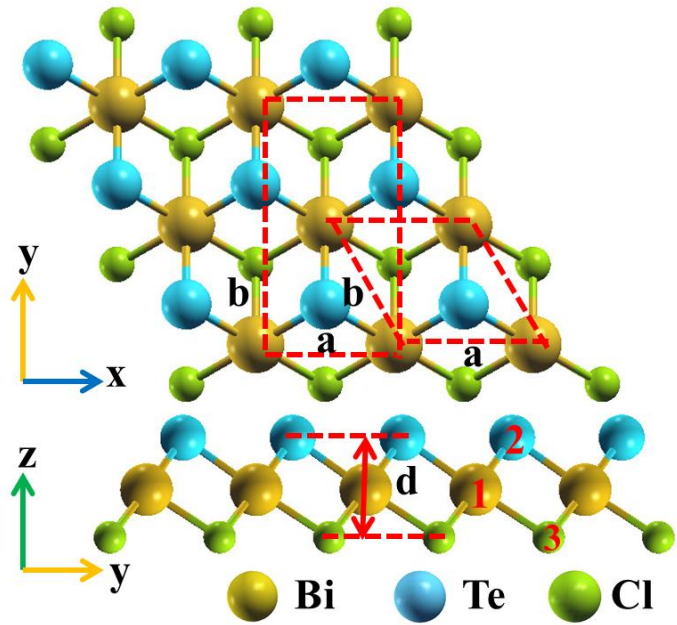

**Figure 1:** Top view and side view of BiTeCl monolayer. Here 1, 2 and 3 represent three atoms in the unit cell. Rectangular and hexagonal unit cell are also shown.

## 3.1 Mechanical properties

The mechanical properties are investigated by applying uniaxial strain in the rectangular unit cell of BiTeCl monolayer (a = 4.39 Å and b = 7.46 Å). We have applied in-plane uniaxial strain ($\varepsilon = \frac{l-l_0}{l_0}$, where $l_0$ and $l$ is the relaxed and deformed lattice constants) along the X and Y directions to calculate the ideal tensile strength and corresponding maximum axial strain. To evaluate the ideal tensile strength, stress-strain relation is needed to be computed which is originally formulated for bulk materials. For 2D crystals, the equivalent stress may be calculated by rescaling the stress with a factor z/d (where z is the length of the cell in the Z direction and d is the thickness of the monolayer) [35]. The ideal tensile strength is the maximum value of stress that can withstand material before the fracture point. The calculated mechanical parameters are presented in Table 1.

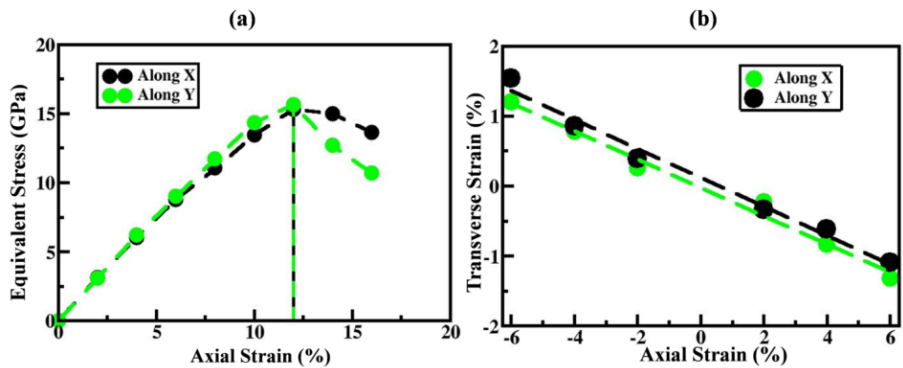

**Figure 2:** **(a)** Axial strain versus equivalent stress curve and **(b)** transverse strain as a function of axial strain for Janus BiTeCl monolayer.

The ideal tensile strength of the Janus BiTeCl monolayer is calculated to be 15.45 GPa and 16.22 GPa at the maximum axial strain of 12% along with X and Y directions, respectively (Figure 2(a)). The ideal strength of the BiTeCl monolayer is higher than that of the BiTeI monolayer (3.46 GPa and 5.97 GPa) [36]. From the stress-strain relation in the harmonic region, Young's modulus is obtained as the ratio of applied stress to the tensile strain ($Y = \sigma/\varepsilon$). The calculated value of Young's modulus for Janus BiTeCl monolayer is 73.58 GPa and 76.7 GPa along X- and Y-direction, respectively. The mechanical properties of the BiTeCl monolayer are calculated to be nearly isotropic (Table 1).

**Table 1:** Ideal tensile strength, maximum axial strain, Young's Modulus (Y), Elastic Modulus and Poisson Ratio (ν) in X and Y directions, respectively for BiTeCl monolayer.

| Direction | Ideal Strength (GPa) | Maximum axial strain (%) | Young's Modulus (GPa) | Elastic Modulus ($Jm^{-2}$) | Poisson Ratio |
|---|---|---|---|---|---|
| X | 15.45 | 12 | 73.58 | $C^{2D}$=29.67 | 0.20 |
| Y | 16.22 | 12 | 76.7 | $C^{2D}$=20.11 | 0.21 |

From the relationship between the axial strain and transverse strain (Figure 2 (b)), we have calculated the Poisson's ratio ($\nu = -d\varepsilon_{transverse}/d\varepsilon_{axial}$). Poisson's ratio is linked with chemical bonding, ductility and brittleness in the material. The calculated value of Poisson's ratio is ~0.2, is in excellent agreement with the previously reported results [19]. The Poisson's ratio value indicates that the chemical bond of this material is majorly the ionic mixing with a slight metallic contribution. Additionally, according to Frantsevich's rule, as the Poisson's ratio value is less than 1/3, the BiTeCl monolayer has a brittle nature [37].

In order to get deep insight of the mechanical behavior of BiTeCl monolayer, we investigated the bonding characteristics as a function of axial strain (Figure 3). On applying strain along X- and Y-directions, a significant change appears in values of bond length and bond angles. The percentage change in bond lengths (R1, R2, and R3) is nearly the same along X and Y-directions at maximum strain (Table S1, ESI) that results in nearly the same ideal strength in BiTeCl monolayer along X and Y directions.

We also calculate the elastic modulus by fitting the strain energy density curve with the applied uniaxial strain (Figure S3(a), ESI). The calculated value of elastic modulus of Janus BiTeCl monolayer is 29.67 Jm$^{-2}$ and 20.11 Jm$^{-2}$ along X- and Y-directions, respectively which is in very good agreement with the recent study [19].

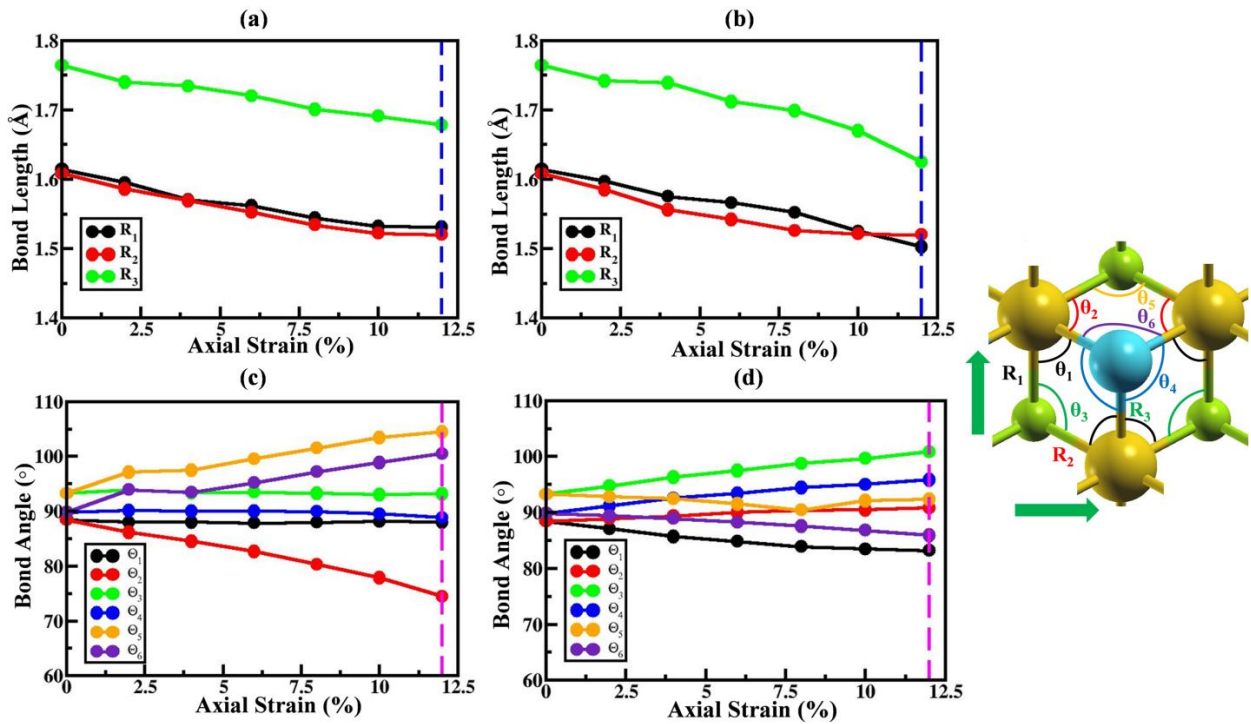

**Figure 3:** Variation of bond length and bond angle of 2D Janus BiTeCl monolayer with axial strain **(a, c)** and **(b, d)** correspond to X- and Y-directions, respectively.

## 3.2 Electronic and Optical Properties

Due to the of heavy elements like Bi and Te and the broken inversion symmetry of the structure, the inclusion of the SOC effects in 2D Janus BiTeCl monolayer leads to the splitting of bands that results in the reduction of bandgap from 1.72 eV at the GGA level of theory (Figure S4, ESI) to 0.84 eV at GGA+SOC level of theory (Figure 4(a)), which is in good agreements with the previously reported value (0.92 eV) [19]. Using hybrid HSE06 functional, the bandgap value without SOC comes out to be 3.06 eV (Figure S5, ESI). The comparison of bandgap value of BiTeCl Janus monolayer using different functional is given in Table S2, ESI. The states near the Fermi level are mainly contributed from the p-orbitals of Cl and Te (Figure 4(b)). The work function of the BiTeCl monolayer is calculated to be 4.38 eV.

We now calculate carrier mobility ($\mu_c$) and relaxation time ($\tau$) using deformation potential theory is given as [38, 39]:

$$\mu_c = \frac{eC^{2D}\hbar^3}{K_B T m m^* E_d^{i^2}} \quad (1)$$

$$\tau = \frac{\mu_c m}{e} \quad (2)$$

where $C^{2D}$ is the elastic modulus, $\tau$ is the relaxation time, m is the effective mass and $E_d^i$ is the deformation potential of Janus BiTeCl monolayer. The shift of band edge that is induced by small strain is given by deformation potential ($E_d^i$) as shown in Figure S3(b), ESI.

**Table 2:** Effective mass (m), deformation potential ($E_d$), carrier mobility ($\mu_c$) and relaxation time ($\tau$) of electrons (e) and holes (h), in X and Y directions, respectively.

| Direction | Effective Mass (m) | Deformation Potential (eV) | Carrier Mobility ($cm^2V^{-1}s^{-1}$) 300K | Relaxation Time (ps) 300K | Relaxation Time (ps) 500 K | Relaxation Time (ps) 700 K |
|---|---|---|---|---|---|---|
| X | $m_e$ = 0.16 | $E_d^e$ = 4.08 | $\mu_e$= 1541.72 | $\tau_e$ = 0.09 | $\tau_e$ = 0.05 | $\tau_e$ = 0.03 |
|   | $m_h$ = 1.53 | $E_d^h$ = 2.50 | $\mu_h$= 39.96 | $\tau_h$ = 0.02 | $\tau_h$ = 0.015 | $\tau_h$ =0.010 |
| Y | $m_e$ = 0.16 | $E_d^e$ = 3.86 | $\mu_e$= 1191.59 | $\tau_e$ = 0.07 | $\tau_e$ = 0.04 | $\tau_e$ = 0.03 |
|   | $m_h$ = 1.82 | $E_d^h$ = 2.6 | $\mu_h$ = 20.99 | $\tau_h$= 0.01 | $\tau_h$ = 0.007 | $\tau_h$ = 0.005 |

Taking the assumption that electronic structure is parabolic, the effective mass m is calculated as m = $\hbar^2$/ [$\partial^2 E(k)/ \partial k^2$], where E(k) is the total energy and k is momentum. The effective mass of electrons and holes is computed by the quadratic fitting of energy band curvature around the VBM and CBM. The calculated value of carrier mobility values shows directional dependence as shown in Table 2. The electron and hole carrier mobility of BiTeCl monolayer is calculated to be higher than BiTeI [36], KAgS [40], KAgSe [40], WS$_2$, WSSe and WSTe [6] monolayers. The relaxation time for electrons and holes shows the isotropic nature (Table 2). Note that, the relaxation time of electrons and holes decreases as the temperature increases.

Next, we have calculated the optical properties of the BiTeCl monolayer. The optical absorbance (Figure 4(c)) of the 2D monolayer is calculated from the imaginary part of dielectric function ($\varepsilon_i(\omega)$) (Figure S6(a), ESI) as [41, 42]:

$$A(\omega) = \frac{\omega}{c} L \varepsilon_i(\omega) \tag{3}$$

where L is the length of the supercell in Z-direction. The peak of the optical absorbance spectrum is obtained at 2.71 eV (in-plane) and 3.63 eV (out-of-plane), respectively (Figure 4(c)), which is comparable with MXY (M = Mo, W and X/Y = S, Se, Te) with prominent absorption peak located at 2.4 – 3.3 eV [11]. The peaks in absorption spectra are closely related to the interband transition in the electronic band structure [43]. These transitions arise from the interband transition from the valance bands below the Fermi level to the conduction bands above the Fermi level. The absorption edge at ~0.8 eV is also consistent with the bandgap calculated from the PBE+SOC level of theory. The static dielectric constants are calculated to be 4.33 and 1.91 from real part of dielectric function (Figure S6(b), ESI) for in-plane and out-of-plane directions, respectively, which are comparable to the dielectric constant values of $WS_2$ (4.4 eV and 2.9 eV) and $WSe_2$ (4.5 eV and 2.9eV) monolayers [44]. Note that the optical absorbance peaks by using GGA level of theory are observed at 3.4 eV and 4.5 eV (Figure S7(a), ESI). To further investigate the optical properties, we have used the GW+BSE method to calculate the exciton binding energy. The optical absorption peaks are shifted to 4.6 eV and 5.2 eV corresponding to in-plane and out-of-plane directions (without SOC effects) as shown in Figure S8(a), ESI. The calculated value of excitonic binding energy is 0.36 eV.

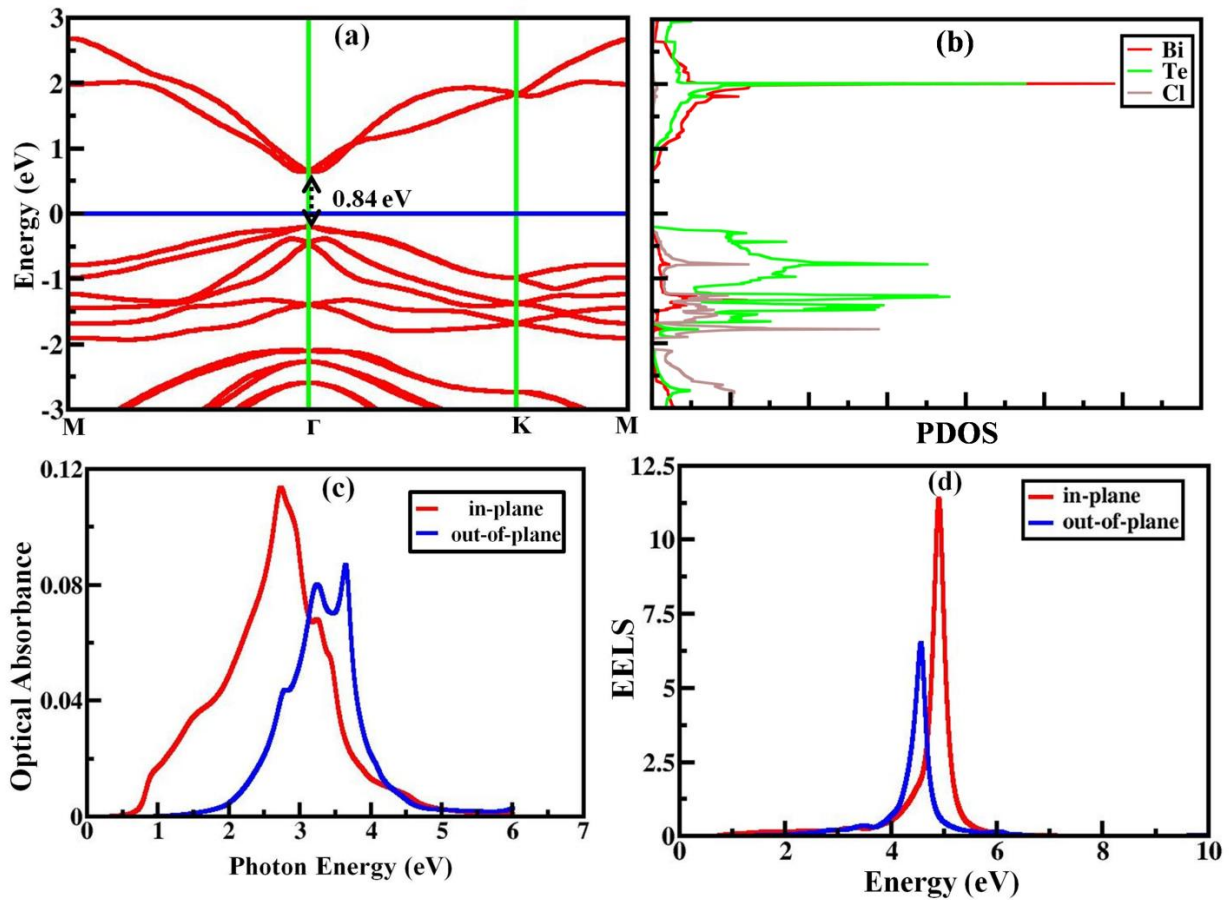

**Figure 4:** **(a)** The electronic band structure, **(b)** Density of states, **(c)** Optical absorbance, and **(d)** Electron energy loss spectra of Janus BiTeCl monolayer at PBE+SOC level of theory.

The electron energy loss spectrum (EELS) of the BiTeCl monolayer is calculated from the imaginary and real part of the dielectric function which is given as [35]:

$$\text{Im}\left\{\frac{-1}{\varepsilon(\omega)}\right\} = \frac{\varepsilon_i(\omega)}{\varepsilon_r^2(\omega) + \varepsilon_i^2(\omega)} \tag{4}$$

where $\varepsilon_r(\omega)$ and $\varepsilon_i(\omega)$ are the real and imaginary parts of the dielectric function. The sharp electron energy loss spectrum peak is obtained at 4.9 eV and 4.5 eV for in-plane and out-of-plane directions, respectively at the GGA+SOC level of theory (Figure 4(d)). Note that the EELS peaks at the GGA level of theory are obtained at 3.4 eV and 2.9 eV for in-plane and out-of-plane directions, respectively (Figure S7(b), ESI). The electron energy loss spectrum peaks shifted to higher value i.e. 8.56 eV and 9.63 eV corresponding to in-plane and out-of-plane direction (without SOC effects) using GW+BSE method (Figure S8(b), ESI). Note that, the $\pi \rightarrow \pi*$ and $\sigma \rightarrow \sigma*$ electrons excitations are reported to occur at low energy (< 5 eV) and high energy (> 10 eV), respectively [44] .The moderate value of bandgap and absorption in the UV-visible region shows that it can be potential material in optoelectronic devices.

## 3.3 Thermoelectric properties

### 3.3.1 Lattice thermal conductivity

The lattice thermal conductivity is calculated using linear Boltzmann transport equation as [45]:

$$\kappa_{l,i} = \sum \sum C_p v_{g,i}^2(\lambda, q) \tau(\lambda, q) \tag{5}$$

where $\kappa_{l,i}$ is the lattice thermal conductivity, $C_p$ is the specific heat capacity, $v_{g,i}$ is the group velocity of phonons, $\lambda$ and $q$ is the phonon mode and wave vector. For lattice thermal conductivity calculation, we have used converged value of k point mesh (Figure S9(a), ESI). The lattice thermal conductivity varies with a temperature of 1/T (Figure 5(a)). The lattice thermal conductivity value of the BiTeCl monolayer is calculated to be 1.46 W/mK at 300 K which is lower than that of bulk BiTeCl (3.1 W/mK) and $Bi_2Te_3$ (2.5 W/mK) [46] and higher than that of KAgX (0.33 W/mK) [25, 40]. The lower value of lattice thermal conductivity of Janus BiTeCl monolayer can be explained in terms of phonon's group velocity ($v_g$) which is estimated as [47]:

$$v_g(\lambda, q) = \frac{\partial \omega(\lambda, q)}{\partial q} \tag{6}$$

where $\omega_{\lambda,q}$ is the phonon frequency. The group velocity value of transverse acoustical (TA) and longitudinal acoustical (LA) mode is calculated to be ~1.5 Km/s and ~1.8 Km/s, respectively (Figure 5(b)). The lower value of group velocity leads to a lower value of lattice thermal conductivity. Further, we quantify the lattice thermal conductivity of Janus BiTeCl monolayer in terms of Gruneisen parameters and phonon relaxation time (Figures 5(c) and 5(d)). A dimensionless quantity i.e. Gruneisen parameter ($\gamma$) describes the anharmonic interactions of crystal structure that is formulated as [48]:

$$\gamma_{\lambda,q} = -\frac{N}{w_{\lambda,q}} \frac{\partial \omega_{\lambda,q}}{\partial N} \tag{7}$$

where N is the volume of the crystal. The higher value of the Gruneisen parameter is related to the large phonon-phonon scattering which results in a lower value of phonon relaxation time which in turn leads to a lower value of lattice thermal conductivity. The value of the lifetime of an acoustic phonon is much higher than that of the optical mode of a phonon (Figure 5(d)). The low value of group velocity, shorter relaxation time and higher Gruneisen parameter possess favorable conditions for phonon transport, resulting extremely low value of lattice thermal conductivity.

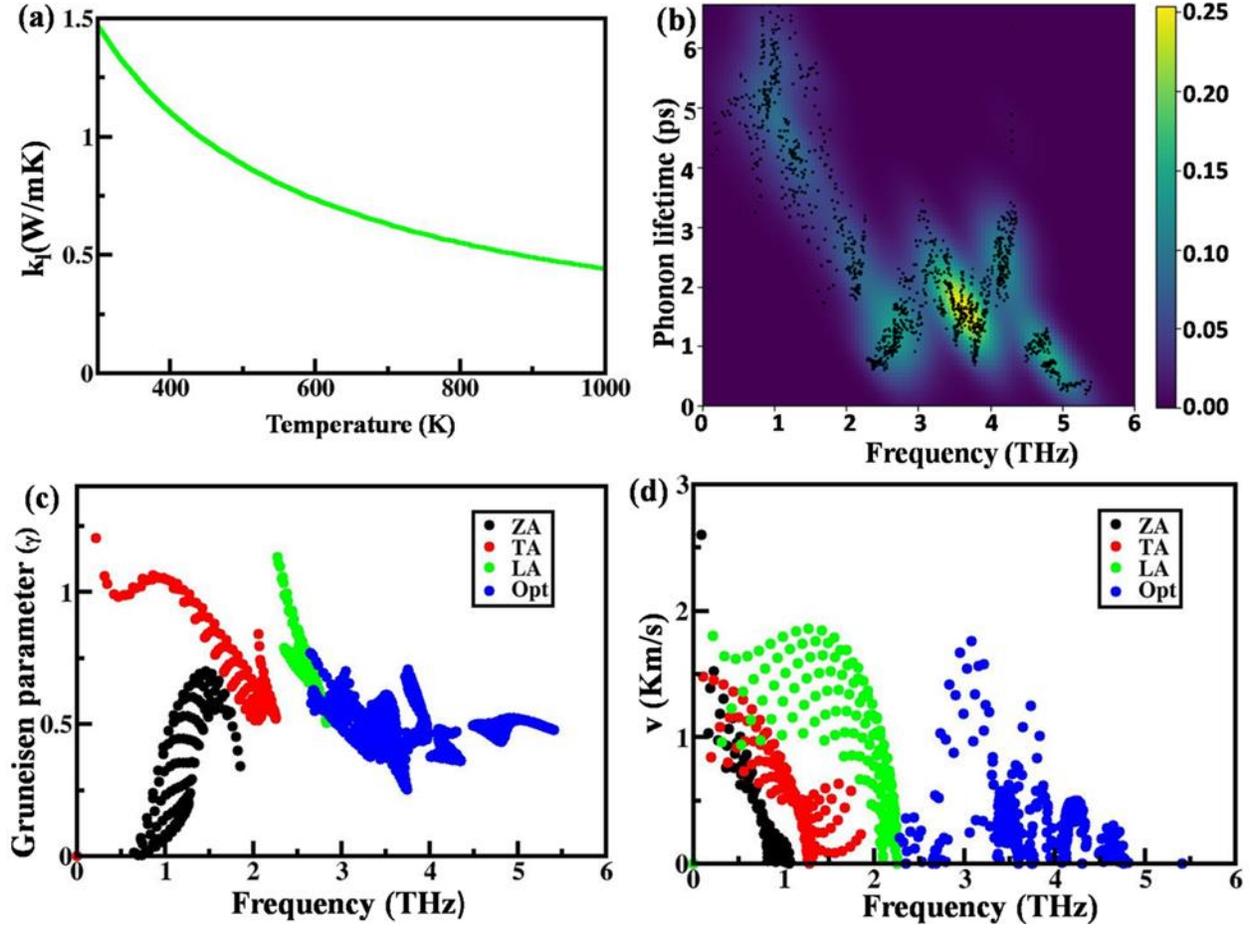

**Figure 5:** **(a)** Lattice thermal conductivity**, (b)** phonon group velocity**, (c)** Gruneisen parameter, and **(d)** Phonon relaxation time of Janus BiTeCl monolayer.

### 3.3.2 Electronic transport properties

Next, we calculate the electronic transport properties using semi-classical Boltzmann transport equation which includes Seebeck coefficient (S), electronic conductivity ($\sigma$), electronic thermal conductivity ($\sigma_e$), and power factor (PF) as [32]:

$$S_{a.b}(T,\mu) = \frac{1}{evT\sigma_{a,b}} \int \Sigma(\varepsilon)(\varepsilon-\mu)\left[\frac{-\partial f_\mu}{\partial \varepsilon}\right] d\varepsilon \qquad (8)$$

$$\sigma_{a,b}(T,\mu) = \frac{1}{V}\int \Sigma(e)\left[\frac{-\partial f_\mu}{\partial \varepsilon}\right] d\varepsilon \qquad (9)$$

$$\Sigma(\varepsilon) = \frac{e^2}{N}\sum \tau \upsilon_a(j,k)\upsilon_b(j,k)\frac{\delta\,(\varepsilon-\varepsilon_{j,k})}{d\varepsilon} \qquad (10)$$

$$\kappa_{a,b}(T,\mu) = \frac{1}{\Omega}\int \sigma_{a,b}\,\varepsilon(\varepsilon-\mu)^2\left[-\frac{\partial f_\mu}{\partial \mu}\right] d\varepsilon \qquad (11)$$

Where, a and b in the above equations represent cartesian indices, $f_\mu$ is Fermi-Dirac distribution function, N is the number of k-points sampling and $\sum(\varepsilon)$ is the transport distribution function. The electronic transport coefficient is calculated at converged value of k-points (Figure S9(b-d), ESI). The maximum value of Seebeck coefficient of other 2D monolayers is also mentioned in literature [49, 50]. Seebeck coefficient of BiTeCl monolayer is calculated to be 2489.25 µV/K, 1584.81 µV/K and 1164.82µV/K at 300K, 500K, and 700 K, respectively (Figure 6 (a)), which is almost the same as calculated through the GGA level of theory (Figure S10(a), ESI). The value of Seebeck coefficient corresponding to the maximum ZT is 278.71 µV/K, 328.90 µV/K and 373.60 µV/K at 300K, 500K and 700K, respectively. Note that the experimentally measured value of Seebeck coefficient for bulk BiTeCl is -218 µV/K (electrons are dominant charge carriers) [46]. The value of the Seebeck coefficient for other related 2D Janus materials such as KAgS, KAgSe, WSSe, WSTe, ReCN and BiTeBr is reported to be 1020 µV/K [40], 998 µV/K [40], 322.26 µV/K [6], 322.15 µV/K [6], 993 µV/K [51] and ~ 200 µV/K [25], respectively at 300 K.

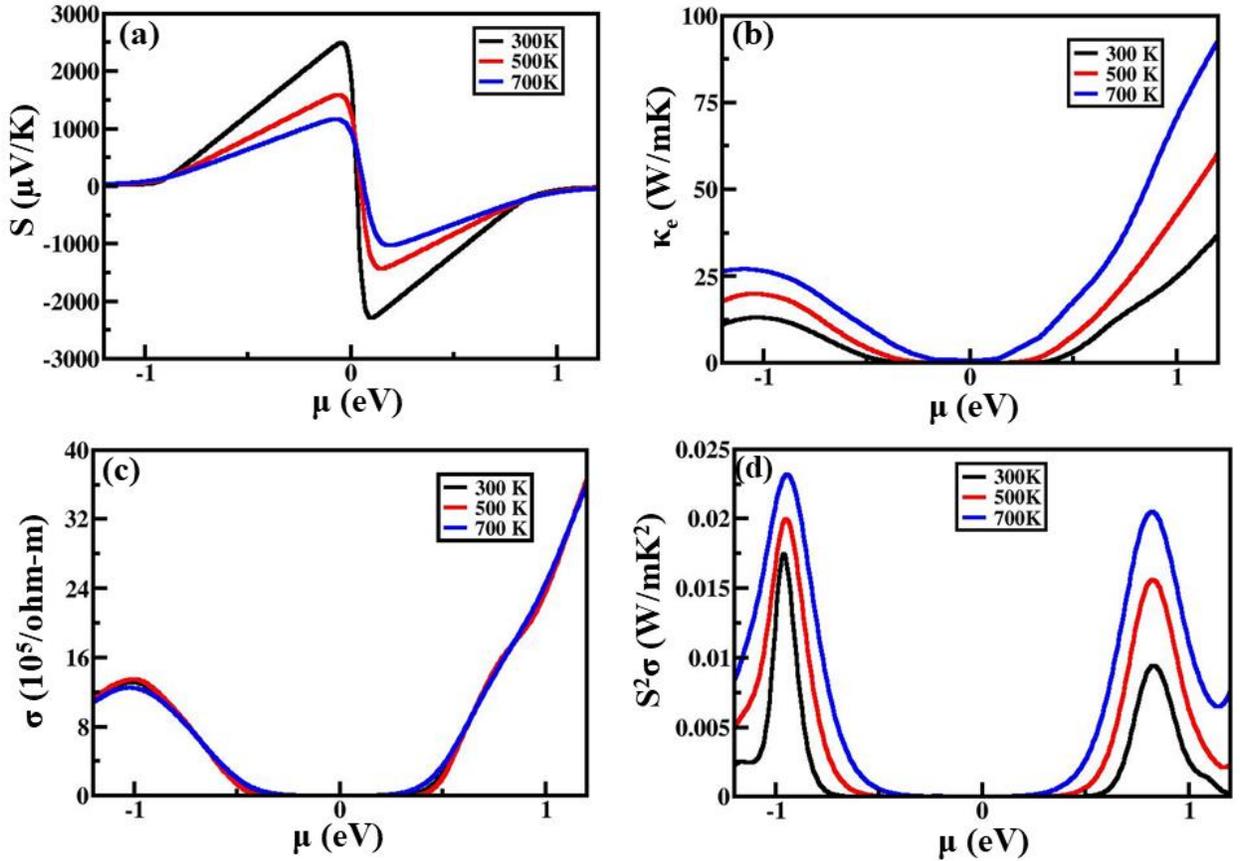

**Figure 6:** **(a)** Seebeck coefficients, **(b)** electrical conductivity, **(c)** electronic lattice thermal conductivity and **(d)** Power factor of Janus BiTeCl monolayer at medium temperature range.

The electronic thermal conductivity of the BiTeCl monolayer is calculated to be ~9.3 (W/mK) at 700 K (Figure 6(b)), which is less than the value obtained without SOC effects i.e. 10.20 W/mK (Figure S10(b), ESI). The electrical conductivity as a function of chemical potential is shown in Figure 6(c). The electrical conductivity is given as:

$$\sigma = ne\mu \tag{12}$$

where n is the charge carrier density. At 700K temperature, the maximum value of electrical conductivity is calculated as 0.83 ($10^5$/ohm-m), which is nearly same to the value ~ 0.84 ($10^5$/ohm-m) from the GGA level of theory (Figure S10(c), ESI). Using the parameter S and σ, the power factor is calculated as:

$$PF = S^2\sigma \tag{13}$$

The power factor value is obtained by the combined effect of the Seebeck coefficient and electrical conductivity. The power factor (PF) value of p-type (n-type) Janus BiTeCl monolayer is ~0.0116 W/mK$^2$ (0.0089 W/mK$^2$) which is smaller than the values ~0.0118 W/mK$^2$ (0.0092 W/mK$^2$) obtained at the GGA level of theory (Figure 6(d) and Figure S10(d), ESI) at 700K.

### 3.3.3 Thermoelectric performance

By the combined effect of power factor and lattice thermal conductivity, we evaluate the figure of merit (ZT) of the Janus BiTeCl monolayer. The thermoelectric figure of merit (ZT) is given as [52, 53]:

$$ZT = \frac{S^2 \sigma T}{K_e + K_l} \tag{14}$$

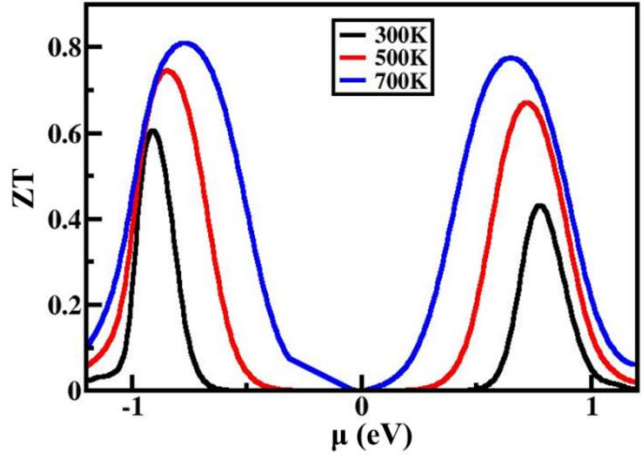

**Figure 7:** Thermoelectric figure of Merit of 2D Janus BiTeCl monolayer.

The calculated value of ZT at 700K using GGA+SOC level of theory (GGA level of theory) is 0.81 (0.79) and 0.77 (0.72) corresponding to p-type and n-type BiTeCl monolayers, respectively, as shown in Figure 7 and Figure S11, ESI. When we include the Lorenz number (L) with different approaches in our calculation, it affects the thermoelectric performance of BiTeCl Janus monolayer (Table S3, ESI). For semiconductors, the fixed value of Lorenz number mentioned in literature is 2.44x10$^{-8}$ W$\Omega$K$^{-2}$ [54, 55]. Using $L = 1.5 + \exp(-\frac{|S|}{116})$, (where S is Seebeck coefficient in µV/K) different value of Lorenz number is obtained corresponding to different value of Seebeck coefficient [56]. The different values of Lorenz number results in the different values of ZT at room temperature. The thermoelectric performance of Janus BiTeCl monolayer in terms of ZT value has been compared with the other Janus monolayers in Table 3, which suggests Janus BiTeCl monolayer to be a good candidate for medium-temperature thermoelectric applications.

**Table 3:** Thermoelectric figure of merit (ZT) of various 2D Janus monolayers.

| 2D materials | ZT | Temperature (in K) | Reference |
|---|---|---|---|
| BiTeCl | 0.43-0.75 | 300-500 | Our study |
| BiTeBr | 0.7-1.75 | 300-600 | [25] |
| WSSe | 0.013 | 300 | [6] |
| WSTe | 0.742 | | |
| KAgS | 2.63-4.05 | 300-500 | [40] |
| KAgSe | 3.11-4.65 | | |
| PdSSe | 0.09-0.33 | 300-600 | [55] |
| PdSTe | 0.26-0.59 | | |
| PdSeTe | 0.98-1.86 | | |
| PtSSe | 0.37-0.64 | | |
| PtSTe | 0.26-0.83 | | |
| PtSeTe | 0.91-1.97 | | |
| HfSSe | 0.86-0.93 | 300-500 | [57] |
| ZrSSe | ~1 | 300-600 | [58] |
| $Ga_2SSe$ | 0.725 | 300 | [59] |
| $In_2SeTe$ | 0.97-1.42 | 300-900 | [60] |
| $Al_2SSe$ | 0.1-0.6 | 300-500 | [61] |
| $In_2SO$ | 0.075-0.5 | 300-900 | [9] |
| $In_2SeO$ | 0.071-0.8 | | |
| $AsSbC_3$ | 0.99-0.95 | 300-700 | [27] |
| AsSBr | 0.21-0.53 | 300-600 | [26] |

**Conclusions**

In summary, mechanical, optical and thermoelectric properties of Janus BiTeCl monolayer are investigated using density functional and Boltzmann transport theory. The mechanical strength of BiTeCl is calculated to be 15.45 GPa and 16.22 GPa along with X and Y directions, respectively. The optical absorption spectrum of the Janus BiTeCl monolayer lies in the UV-visible region. The electron energy loss spectra are observed at 4.9 eV and 4.5 eV for in-plane and out-of-plane directions, respectively. The calculated value of the Seebeck coefficient and electrical conductivity is 328.90 µV/K and 1.20 ($10^5$/ohm-m) at 500 K. The higher value of the Gruneisen parameter, lower value of group velocity and phonon relaxation time results in an extremely low value of lattice thermal conductivity at room temperature. By the combined effect of electronic transport coefficient and lattice thermal conductivity, the calculated ZT value 0.75 (p-type) and 0.67 (n-type) at 500 K shows the good thermoelectric performance of the Janus BiTeCl monolayer.

## Acknowledgments

PC and JS acknowledge the financial assistance from CSIR in the form of junior research fellowship (JRF) and senior research fellowship (SRF), respectively. The computational facilities at the Department of Physics at the Central University of Punjab are used to obtain the results presented in the paper. Helpful discussion with Mukesh Jakhar is highly acknowledged.

## References

[1] L. Zhang, Z. Yang, T. Gong, R. Pan, H. Wang, Z. Guo, H. Zhang, X. Fu, Recent advances in emerging Janus two-dimensional materials: from fundamental physics to device applications, Journal of Materials Chemistry A, 8 (2020) 8813-8830.
[2] J. Zhang, S. Jia, I. Kholmanov, L. Dong, D. Er, W. Chen, H. Guo, Z. Jin, V.B. Shenoy, L. Shi, Janus monolayer transition-metal dichalcogenides, ACS nano, 11 (2017) 8192-8198.
[3] A.-Y. Lu, H. Zhu, J. Xiao, C.-P. Chuu, Y. Han, M.-H. Chiu, C.-C. Cheng, C.-W. Yang, K.-H. Wei, Y. Yang, Janus monolayers of transition metal dichalcogenides, Nature nanotechnology, 12 (2017) 744-749.
[4] V. Van Thanh, N.D. Van, R. Saito, N.T. Hung, First-principles study of mechanical, electronic and optical properties of Janus structure in transition metal dichalcogenides, Applied Surface Science, 526 (2020) 146730.
[5] C. Xia, W. Xiong, J. Du, T. Wang, Y. Peng, J. Li, Universality of electronic characteristics and photocatalyst applications in the two-dimensional Janus transition metal dichalcogenides, Physical Review B, 98 (2018) 165424.
[6] A. Patel, D. Singh, Y. Sonvane, P. Thakor, R. Ahuja, High Thermoelectric Performance in Two-Dimensional Janus Monolayer Material WS-X (X= Se and Te), ACS applied materials & interfaces, 12 (2020) 46212-46219.
[7] Y.-D. Guo, H.-B. Zhang, H.-L. Zeng, H.-X. Da, X.-H. Yan, W.-Y. Liu, X.-Y. Mou, A progressive metal–semiconductor transition in two-faced Janus monolayer transition-metal chalcogenides, Physical Chemistry Chemical Physics, 20 (2018) 21113-21118.
[8] M. Demirtas, M.J. Varjovi, M.M. Çiçek, E. Durgun, Tuning structural and electronic properties of two-dimensional aluminum monochalcogenides: Prediction of Janus Al 2 X X'(X/X': O, S, Se, Te) monolayers, Physical Review Materials, 4 (2020) 114003.

[9] T.V. Vu, C.V. Nguyen, H.V. Phuc, A. Lavrentyev, O. Khyzhun, N.V. Hieu, M. Obeid, D. Rai, H.D. Tong, N.N. Hieu, Theoretical prediction of electronic, transport, optical, and thermoelectric properties of Janus monolayers In 2 X O (X= S, Se, Te), Physical Review B, 103 (2021) 085422.

[10] H. Yang, P. Zhao, Y. Ma, X. Lv, B. Huang, Y. Dai, Janus single-layer group-III monochalcogenides: a promising visible-light photocatalyst, Journal of Physics D: Applied Physics, 52 (2019) 455303.

[11] J. Wang, H. Shu, T. Zhao, P. Liang, N. Wang, D. Cao, X. Chen, Intriguing electronic and optical properties of two-dimensional Janus transition metal dichalcogenides, Physical Chemistry Chemical Physics, 20 (2018) 18571-18578.

[12] T.V. Vu, H.D. Tong, D.P. Tran, N.T. Binh, C.V. Nguyen, H.V. Phuc, H.M. Do, N.N. Hieu, Electronic and optical properties of Janus ZrSSe by density functional theory, RSC Advances, 9 (2019) 41058-41065.

[13] S. Ahmad, F. Khan, B. Amin, I. Ahmad, Effect of strain on structural and electronic properties, and thermoelectric response of MXY (M= Zr, Hf and Pt; X/Y= S, Se) vdW heterostructures; A first principles study, Journal of Solid State Chemistry, 299 (2021) 122189.

[14] S.-D. Guo, X.-S. Guo, Y. Deng, Tuning the electronic structures and transport coefficients of Janus PtSSe monolayer with biaxial strain, Journal of Applied Physics, 126 (2019) 154301.

[15] X. Yang, D. Singh, Z. Xu, Z. Wang, R. Ahuja, An emerging Janus MoSeTe material for potential applications in optoelectronic devices, Journal of Materials Chemistry C, 7 (2019) 12312-12320.

[16] W. Shi, Z. Wang, Mechanical and electronic properties of Janus monolayer transition metal dichalcogenides, Journal of Physics: Condensed Matter, 30 (2018) 215301.

[17] D. Hoat, M. Naseri, N.N. Hieu, R. Ponce-Pérez, J. Rivas-Silva, T.V. Vu, G.H. Cocoletzi, A comprehensive investigation on electronic structure, optical and thermoelectric properties of the HfSSe Janus monolayer, Journal of Physics and Chemistry of Solids, 144 (2020) 109490.

[18] D. Hajra, R. Sailus, M. Blei, K. Yumigeta, Y. Shen, S. Tongay, Epitaxial Synthesis of Highly Oriented 2D Janus Rashba Semiconductor BiTeCl and BiTeBr Layers, ACS nano, 14 (2020) 15626-15632.

[19] M.J. Varjovi, E. Durgun, First-principles study on structural, vibrational, elastic, piezoelectric, and electronic properties of the Janus Bi X Y (X= S, Se, Te and Y= F, Cl, Br, I) monolayers, Physical Review Materials, 5 (2021) 104001.

[20] I. Waller, Dynamical theory of crystal lattices by M. Born and K. Huang, Acta Crystallographica, 9 (1956) 837-838.

[21] M. Samanta, T. Ghosh, S. Chandra, K. Biswas, Layered materials with 2D connectivity for thermoelectric energy conversion, Journal of Materials Chemistry A, 8 (2020) 12226-12261.

[22] Z. Feng, Y. Fu, Y. Zhang, D.J. Singh, Characterization of rattling in relation to thermal conductivity: Ordered half-Heusler semiconductors, Physical Review B, 101 (2020) 064301.

[23] M. Zare, B.Z. Rameshti, F.G. Ghamsari, R. Asgari, Thermoelectric transport in monolayer phosphorene, Physical Review B, 95 (2017) 045422.

[24] G. Ding, Y. Hu, D. Li, X. Wang, A comparative study of thermoelectric properties between bulk and monolayer SnSe, Results in Physics, 15 (2019) 102631.

[25] S.-D. Guo, H.-C. Li, Monolayer enhanced thermoelectric properties compared with bulk for BiTeBr, Computational Materials Science, 139 (2017) 361-367.

[26] M. Liu, S.-B. Chen, C.-E. Hu, Y. Cheng, H.-Y. Geng, Thermoelectric properties of Janus AsSBr monolayer from first-principles study, Solid State Communications, 342 (2022) 114612.

[27] A. Marjaoui, M.A. Tamerd, A. El Kasmi, M. Diani, M. Zanouni, First-principles study on electronic and thermoelectric properties of Janus monolayers AsXC3 (X: Sb, Bi), Computational Condensed Matter, (2021) e00623.

[28] P. Giannozzi, S. Baroni, N. Bonini, M. Calandra, R. Car, C. Cavazzoni, D. Ceresoli, G.L. Chiarotti, M. Cococcioni, I. Dabo, QUANTUM ESPRESSO: a modular and open-source software project for quantum simulations of materials, Journal of physics: Condensed matter, 21 (2009) 395502.

[29] P. Giannozzi, O. Andreussi, T. Brumme, O. Bunau, M.B. Nardelli, M. Calandra, R. Car, C. Cavazzoni, D. Ceresoli, M. Cococcioni, Advanced capabilities for materials modelling with Quantum ESPRESSO, Journal of physics: Condensed matter, 29 (2017) 465901.
[30] J.P. Perdew, K. Burke, M. Ernzerhof, Generalized gradient approximation made simple, Physical review letters, 77 (1996) 3865.
[31] A. Marini, C. Hogan, M. Grüning, D. Varsano, Yambo: an ab initio tool for excited state calculations, Computer Physics Communications, 180 (2009) 1392-1403.
[32] G.K. Madsen, D.J. Singh, BoltzTraP. A code for calculating band-structure dependent quantities, Computer Physics Communications, 175 (2006) 67-71.
[33] A. Togo, L. Chaput, I. Tanaka, Distributions of phonon lifetimes in Brillouin zones, Physical review B, 91 (2015) 094306.
[34] A. Bafekry, S. Karbasizadeh, C. Stampfl, M. Faraji, H. Do Minh, A.S. Sarsari, S. Feghhi, M. Ghergherehchi, Two-dimensional Janus Semiconductors BiTeCl and BiTeBr monolayers: A first-principles study of the tunable electronic properties via electric field and mechanical strain, Physical Chemistry Chemical Physics, (2021).
[35] J. Singh, P. Jamdagni, M. Jakhar, A. Kumar, Stability, electronic and mechanical properties of chalcogen (Se and Te) monolayers, Physical Chemistry Chemical Physics, 22 (2020) 5749-5755.
[36] W.-Z. Xiao, H.-J. Luo, L. Xu, Elasticity, piezoelectricity, and mobility in two-dimensional BiTeI from a first-principles study, Journal of Physics D: Applied Physics, 53 (2020) 245301.
[37] I. Frantsevich, Elastic constants and elastic moduli of metals and insulators, Reference book, (1982).
[38] J. Bardeen, W. Shockley, Deformation potentials and mobilities in non-polar crystals, Physical review, 80 (1950) 72.
[39] Y. Cai, G. Zhang, Y.-W. Zhang, Polarity-reversed robust carrier mobility in monolayer $MoS_2$ nanoribbons, Journal of the American Chemical Society, 136 (2014) 6269-6275.
[40] X.-L. Zhu, H. Yang, W.-X. Zhou, B. Wang, N. Xu, G. Xie, KAgX (X= S, Se): High-Performance Layered Thermoelectric Materials for Medium-Temperature Applications, ACS Applied Materials & Interfaces, 12 (2020) 36102-36109.
[41] A. Kumar, G. Sachdeva, R. Pandey, S.P. Karna, Optical absorbance in multilayer two-dimensional materials: Graphene and antimonene, Applied Physics Letters, 116 (2020) 263102.
[42] L. Matthes, P. Gori, O. Pulci, F. Bechstedt, Universal infrared absorbance of two-dimensional honeycomb group-IV crystals, Physical Review B, 87 (2013) 035438.
[43] A. Kumar, P. Ahluwalia, A first principle comparative study of electronic and optical properties of $1H-MoS_2$ and $2H-MoS_2$, Materials Chemistry and Physics, 135 (2012) 755-761.
[44] A. Kumar, P. Ahluwalia, Tunable dielectric response of transition metals dichalcogenides $MX_2$ (M= Mo, W; X= S, Se, Te): Effect of quantum confinement, Physica B: Condensed Matter, 407 (2012) 4627-4634.
[45] W.X. Zhou, Y. Cheng, K.Q. Chen, G. Xie, T. Wang, G. Zhang, Thermal conductivity of amorphous materials, Advanced Functional Materials, 30 (2020) 1903829.
[46] J. Jacimovic, X. Mettan, A. Pisoni, R. Gaal, S. Katrych, L. Demko, A. Akrap, L. Forró, H. Berger, P. Bugnon, Enhanced low-temperature thermoelectrical properties of BiTeCl grown by topotactic method, Scripta Materialia, 76 (2014) 69-72.
[47] S.-i. Tamura, Y. Tanaka, H.J. Maris, Phonon group velocity and thermal conduction in superlattices, Physical Review B, 60 (1999) 2627.
[48] D. Broido, A. Ward, N. Mingo, Lattice thermal conductivity of silicon from empirical interatomic potentials, Physical Review B, 72 (2005) 014308.
[49] J. Gu, X. Qu, Excellent thermoelectric properties of monolayer RbAgM (M= Se and Te): first-principles calculations, Physical Chemistry Chemical Physics, 22 (2020) 26364-26371.

[50] S. Ahmadi, M. Raeisi, L. Eslami, A. Rajabpour, Thermoelectric Characteristics of Two-Dimensional Structures for Three Different Lattice Compounds of B–C–N and Graphene Counterpart BX (X= P, As, and Sb) Systems, The Journal of Physical Chemistry C, 125 (2021) 14525-14537.

[51] A.M. Reyes, J.L. Ponce-Ruiz, E.S. Hernández, A.R. Serrato, Novel Thermoelectric Character of Rhenium Carbonitride, ReCN, ACS omega, 6 (2021) 18364-18369.

[52] J. Yang, L. Xi, W. Qiu, L. Wu, X. Shi, L. Chen, J. Yang, W. Zhang, C. Uher, D.J. Singh, On the tuning of electrical and thermal transport in thermoelectrics: an integrated theory–experiment perspective, Npj Computational Materials, 2 (2016) 1-17.

[53] W.G. Zeier, A. Zevalkink, Z.M. Gibbs, G. Hautier, M.G. Kanatzidis, G.J. Snyder, Thinking like a chemist: intuition in thermoelectric materials, Angewandte Chemie International Edition, 55 (2016) 6826-6841.

[54] M. Thesberg, H. Kosina, N. Neophytou, On the Lorenz number of multiband materials, Physical Review B, 95 (2017) 125206.

[55] W.-L. Tao, J.-Q. Lan, C.-E. Hu, Y. Cheng, J. Zhu, H.-Y. Geng, Thermoelectric properties of Janus MXY (M= Pd, Pt; X, Y= S, Se, Te) transition-metal dichalcogenide monolayers from first principles, Journal of Applied Physics, 127 (2020) 035101.

[56] H.-S. Kim, Z.M. Gibbs, Y. Tang, H. Wang, G.J. Snyder, Characterization of Lorenz number with Seebeck coefficient measurement, APL materials, 3 (2015) 041506.

[57] J. Bera, A. Betal, S. Sahu, Ultralow lattice thermal conductivity and high thermoelectric performance near room temperature of Janus monolayer HfSSe, arXiv preprint arXiv:2003.02439, (2020).

[58] S.-D. Guo, Y.-F. Li, X.-S. Guo, Predicted Janus monolayer ZrSSe with enhanced n-type thermoelectric properties compared with monolayer ZrS2, Computational Materials Science, 161 (2019) 16-23.

[59] H.T. Nguyen, V.T. Vi, T.V. Vu, N.V. Hieu, D.V. Lu, D. Rai, N.T. Binh, Spin–orbit coupling effect on electronic, optical, and thermoelectric properties of Janus Ga 2 SSe, RSC Advances, 10 (2020) 44785-44792.

[60] A. Marjaoui, M. Zanouni, M. Ait Tamerd, A. El Kasmi, M. Diani, A First-Principles Investigation on Electronic Structure, Optical and Thermoelectric Properties of Janus In2SeTe Monolayer, Journal of Superconductivity and Novel Magnetism, (2021) 1-12.

[61] G.S. Khosa, S. Gupta, R. Kumar, First-principles investigations of electronic and thermoelectric properties of Janus Al2SSe monolayer, Physica B: Condensed Matter, 615 (2021) 413057.